\documentclass[11pt]{article}
\usepackage{abstract}
\usepackage{makecell}
\usepackage{array}
\usepackage{wrapfig}
\usepackage[noadjust]{cite}
\usepackage[utf8]{inputenc}

\usepackage{authblk}
\usepackage{amsmath}
\usepackage{blindtext}
\usepackage{hyperref}

\usepackage{IEEEtrantools}
\usepackage{geometry}
 \usepackage{booktabs}
\usepackage{amsfonts} 
\usepackage{amssymb}
\usepackage{float}
\usepackage[nottoc]{tocbibind}
\usepackage{graphicx}
\graphicspath{{./image/}}
\usepackage{mathtools}

\usepackage{multicol}
\font\myfont=cmr12 at 12pt
\title{{\textbf{A note on the Sine-Gordon expansion method and its applications}}}
\author{\large \textbf{Nizhum Rahman}\\ Email: \href{mailto:nizhum.rahman@uq.edu.au}{nizhum.rahman@uq.edu.au} }

\affil{\myfont School of Mathematics and Physics, The University of Queensland, St Lucia, 4072, Australia \vspace{-2ex}}

\affil{\myfont Faculty of Science and Information Technology, Daffodil International University Dhaka,1207, Bangladesh \vspace{-2ex}}
 
\date{\today}

\begin{document}

\maketitle

\section*{Abstract}
The sine-Gordon expansion method; which is a transformation of the sine-Gordon equation has been applied to the potential-YTSF equation of dimension (3+1) and the reaction-diffusion equation. We obtain new solitons of this equation in the form hyperbolic, complex and trigonometric function by using this method. We plot 2D and 3D graphics of these solutions using symbolic software.  
\newline
\newline \textbf{Keywords:} Sine-Gordon expansion method; the  potential-YTSF equation of dimension (3+1); the reaction-diffusion equation; trigonometric function solutions; hyperbolic function solutions;  complex solutions.
\section{Introduction}
In recent decades, the nonlinear evolution equations (NLEEs) have become one of the vital topics of interest among the researchers from different field including physics, mathematics, engineering and biology \cite{a1,a2,a3,a4,a5,a6,a7}. The analytic solutions in the form of a travelling wave of the NLEEs play a crucial role for the development of scientific fields. A variety of effective methods have been constructed from the end of the $20^{th}$ century by the researchers. Many of these methods have been used successfully in different NLEEs to obtain new solutions \cite{a13,a10,a11,a8,a14,a15,a16,a17,a18,a42,a19,a43,a20,a41,a21,a22,a23,a26,a27,a46,a28,a44,a40,a45,a47}.\\
In this article we will apply the sine-Gordon expansion method  \cite{a9} to the potential-YTSF equation of dimension (3+1) and the reaction-diffusion equation. Recently this method has been applied to only a few research articles \cite{a31,a32,a33,a39}. The remaining part of the article is described as follows. The sine-Gordon expansion method has been stated in section 2. The next section is the applications of this method. Finally, we describe the necessity of this method briefly in the last section. 

\section{Method}
The sine-Gordon equation (SGE) \cite{a9}
\begin{equation}\label{eq1}
 \frac{\partial^2 u}{\partial x^2}-\frac{\partial^2 u}{\partial t^2}=h^2 sin(u)   
\end{equation}
arises in many physical science applications \cite{a51,a52}, where $h$ is a constant. By considering the moving coordinate as $u(x,t)=U(\eta)$ where $\eta=k(x-ct)$, after simplifying this equation, we get
\begin{equation}\label{eq2}
 \frac{d^2 U}{d \eta^2}=\frac{h^2}{k^2(1-c^2)}sin(U)  
\end{equation}
where $\eta$ and $c$ represent the width and velocity of the travelling waves respectively. After multiplying $\frac{d U}{d\eta}$ on both sides of Eq. \eqref{eq2} and integrating, we get,
\begin{equation}\label{eq3}
    \Big[\Big(\frac{U}{2}\Big)'\Big]^2=\frac{h^2}{k^2(1-c^2)}sin^2 \Big(\frac{U}{2}\Big), 
\end{equation}
where the constant of integration is considered to zero.
Substituting $\omega(\eta)=\frac{U}{2}$ and $\frac{h^2}{k^2(1-c^2)}=1$ into Eq.\eqref{eq3}, we obtain
\begin{equation}\label{eq4}
 \omega'=sin\,(\omega).  
\end{equation}
This is a modified formation of the SGE. We can write the solution of Eq. \eqref{eq4} as of the form
\begin{equation}\label{eq5}
  sin\, [\omega(\eta)]= \frac{2 r\,  exp(\eta)}{r^2\, exp(2\eta)+1}\bigg|_{r=1} =sech(\eta),
\end{equation} or
\begin{equation}\label{eq6}
  cos\, [\omega(\eta)]= \frac{r^2 \,exp(2\eta)-1}{r^2\, exp(2\eta)+1}\bigg|_{r=1} =tanh(\eta), 
\end{equation}
with $r$ is a non-zero constant of integration.
\par The travelling wave solution $U(\eta)$  of the NLEEs of the form
\begin{equation}\label{pde}
    Q(u,u_x,u_t,u_{xx},u_{tx},u_{tt},.............)=0,
\end{equation}
can be written as
\begin{equation}\label{eq7}
    U(\eta)=\sum_{j=1}^{n} tanh^{j-1}(\eta)\,[B_j\, sech (\eta)+A_j\, tanh(\eta)]+A_0.
\end{equation}
Using Eq. \eqref{eq5} and Eq. \eqref{eq6}, the solutions of the Eq. \eqref{eq7} take the following form
\begin{equation}\label{eq8}
    U(\omega)=\sum_{j=1}^{n} cos^{j-1}(\omega)\,[B_j\, sin (\omega)+A_j\, cos(\omega)]+A_0.
\end{equation}
We calculate the value of unknown $n$ by apply the homogeneous balance assumption. After letting the collection of
the coefficients of $sin^i\,(\omega) \,cos^j\,(\omega) $
 of equal power to be all $0$ (zero), we get an algebraic system of equations. By find out the unknowns of the system and using 
Eq.\eqref{eq8}, we
get the solutions to Eq. \eqref{pde} is the form of \eqref{eq7}.
\section{Applications}
We can implement the sine-Gordon expansion method in many NLEEs. To demonstrate our method, we examine the method to the potential-YTSF equation of dimension (3+1) and the reaction-diffusion equation.
\subsection{The  potential-YTSF equation of dimension (3+1)}
We consider the  potential-YTSF equation of dimension (3+1) \cite{a34,a53,a54} in the form
\begin{equation}\label{eq10}
  -4\,u_{xt} + u_{xxxz}+4\,u_{x}\,u_{xz} +2\, u_{xx} u_{z} + 3\,u_{yy}  = 0,  
\end{equation}\label{eq11}
which arise in many physical problems and have been solved in different ways by the researchers\cite{a35,a36,a37}. The moving coordinate 
\begin{equation}
    u(x,y,z,t)=U(\eta)\,,\,  \, \eta=x+y+z-ct,
\end{equation}
allows to convert Eq.\eqref{eq10} in to an ODE
\begin{equation*}
    U''''+6U'U''+(4c+3)U''=0, 
\end{equation*}
integrating it with respect to $\eta$, we acquire 
\begin{equation}\label{eq12}
  U'''+3(U')^2+(4c+3)U'=0,  
\end{equation}
where integrating constant is set to zero. Setting $U'=V$, we have
\begin{equation}\label{eq13}
V''+3V^2+(4c+3)V=0.   
\end{equation}
Applying homogeneous balance principle between $V''$ and $V^2$ in Eq. \eqref{eq13} based on Eq. \eqref{eq8} we obtain $n+2=n+n$, which implies $n=2$. 
Now we can write Eq. \eqref{eq8} as
\begin{equation}\label{eq14}
     V(\omega)= B_1\, sin (\omega)+A_1\, cos(\omega)+B_2\, sin (\omega)\,cos(\omega) +A_2\, cos^2(\omega)+A_0
\end{equation}
and 
\begin{IEEEeqnarray}{rCl}\label{eq15}
V''&=&-B_1\,sin^3(\omega)+B_1\,cos^2(\omega)\,sin(\omega)-2A_1\,sin^2(\omega)\,cos(\omega)-5\,B_2\,cos(\omega)\,sin^3(\omega)+B_2\,cos^3(\omega)\,sin(\omega)\nonumber\\&&
  +\>2\,A_2\,sin^4(\omega)-4\,A_2cos^2(\omega)\,sin^2(\omega).
\end{IEEEeqnarray}
By substituting Eq. \eqref{eq14} and Eq. \eqref{eq15} into Eq. \eqref{eq13} and equaling all polynomials with same degree to zero, we get a system of equation as below.\newline
\newline
constant: \,\,\,\,\,\,\,\,\,\,\,\,\,\,$4\,v\,A_{{0}}+3\,{A_{{0}}}^{2}+3
\,{B_{{1}}}^{2}+3\,A_{{0}}+2\,A_{{2}}
=0$\\
$cos(\omega):\,\,\,\,\,\,\,\,\,\,\,\,\,\,\,\,\,\,\, 4\,v\,A_{{1}}+6\,A_{{0}}A_{{1
}}+6\,B_{{1}}B_{{2}}+A_{{1}}
=0$\\
$sin(\omega):  \,\,\,\,\,\,\,\,\,\,\,\,\,\,\,\,\, \,\,       4\,v\,B_{{1}}+6\,A_{{0}}B_{{1}}+2\,B_{{1}}=0$\\
$cos^2(\omega):\,\,\,\,\,\,\,\,\,\,\,\,\,\,\,\,\,4\,
v\,A_{{2}}+6\,A_{{0}}A_{{2}}+3\,{A_{{1}}}^{2}-3\,{B_{{1}}}^{2}+3\,{B_{{2
}}}^{2}-5\,A_{{2}}
=0$\\
$sin(\omega)\,cos(\omega):\,\,4\,v\,B_
{{2}}+6\,A_{{0}}B_{{2}}+6\,A_{{1}}B_{{1}}-2\,B_{{2}}
0=0$\\
$cos^3(\omega):\,\,\,\,\,\,\,\,\,\,\,\,\,\,\,\,\,\, 6\,A
_{{1}}A_{{2}}-6\,B_{{1}}B_{{2}}+2\,A_{{1}}
=0$\\
$sin(\omega)\,cos^2(\omega):6\,A_{{1
}}B_{{2}}+6\,A_{{2}}B_{{1}}+2\,B_{{1}}
=0$\\
$cos^4(\omega):\,\,\,\,\,\,\,\,\,\,\,\,\,\,\,\,\,\, 3\,
{A_{{2}}}^{2}-3\,{B_{{2}}}^{2}+6\,A_{{2}}
=0$\\
$sin(\omega)\,cos^3(\omega):6\,A_{{2}}B_{{2}}+6\,B_{{2}}=0$
\newline
\par
After solving the above system , we find the traveling wave solution $U(\eta)$ to Eq. \eqref{eq10}  in the form of \eqref{eq7}.
\par
Case-1
\begin{equation*}
    c=-7/4\,,\,A_0=2\,,\,A_1=0\,,\,B_1=0\,,\,A_2=-2 \,,\,B_2=0\,, 
    \end{equation*}
which gives:

\begin{equation}\label{sol1}
u(x,y,z,t)=\,U(\eta)=\,2\,tanh\,(\eta), \text{with}\,\, \eta =x+y+z+\frac{7}{4}\, t .
\end{equation}
\begin{figure}[H]
\centering
    \includegraphics[trim={0.5cm 11cm 0cm 10cm},clip=true,totalheight=.25\textheight ]{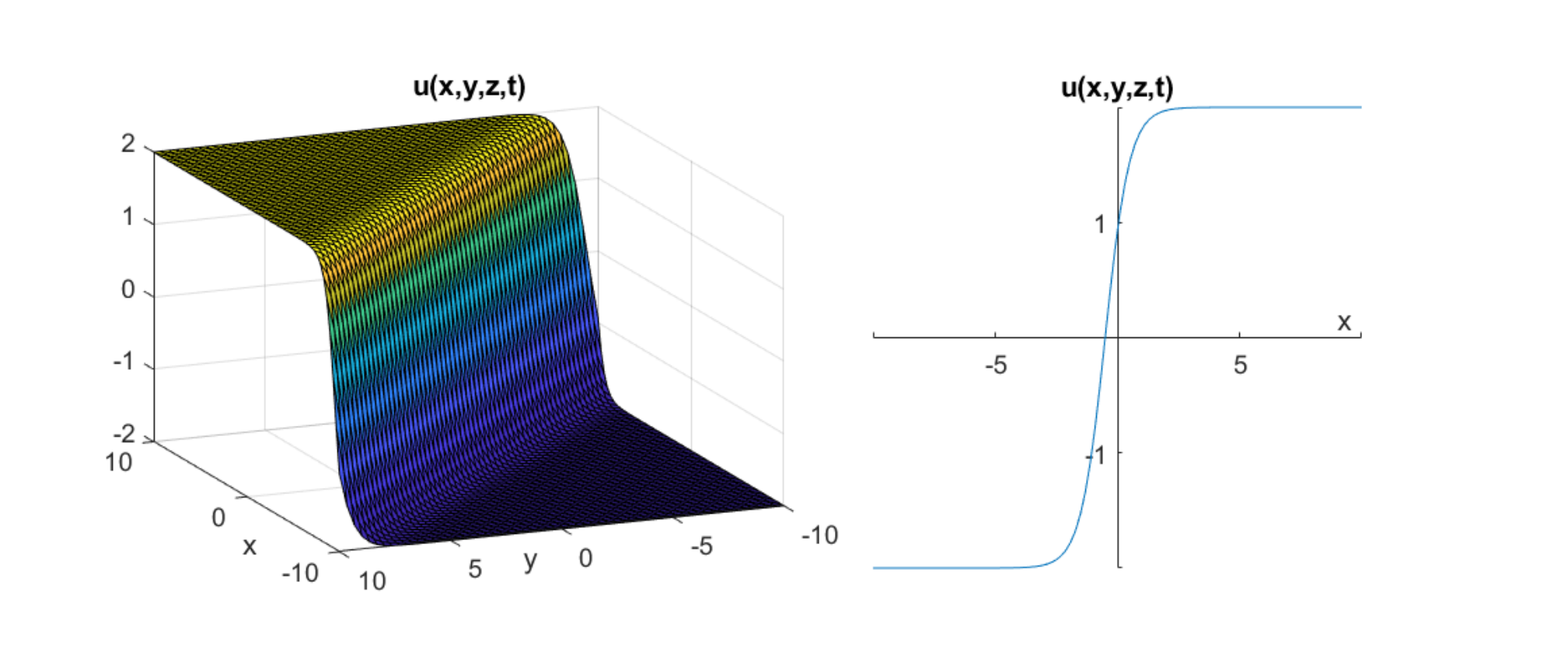}
    \caption{The graphical representation of Eq. \eqref{sol1} (3D on the left side and 2D on the right side).}
    \label{fig1}
\end{figure}
Case-2
\begin{equation*}
    c=1/4\,,\,A_0=2/3\,,\,A_1=0\,,\,B_1=0\,,\,A_2=-2 \,,\,B_2=0\,, 
    \end{equation*}
which gives:

\begin{equation}\label{sol2}
u(x,y,z,t)\,= U(\eta)=\,2\,tanh\,(\eta) -\frac{4}{3}\, \eta, \, \text{with}\,\,\eta=\, x+y+z-\frac{1}{4}\, t.
\end{equation}
\begin{figure}[H]
\centering
    \includegraphics[trim={0.5cm 11.58cm 0cm 10.7cm},clip=true,totalheight=.25\textheight ]{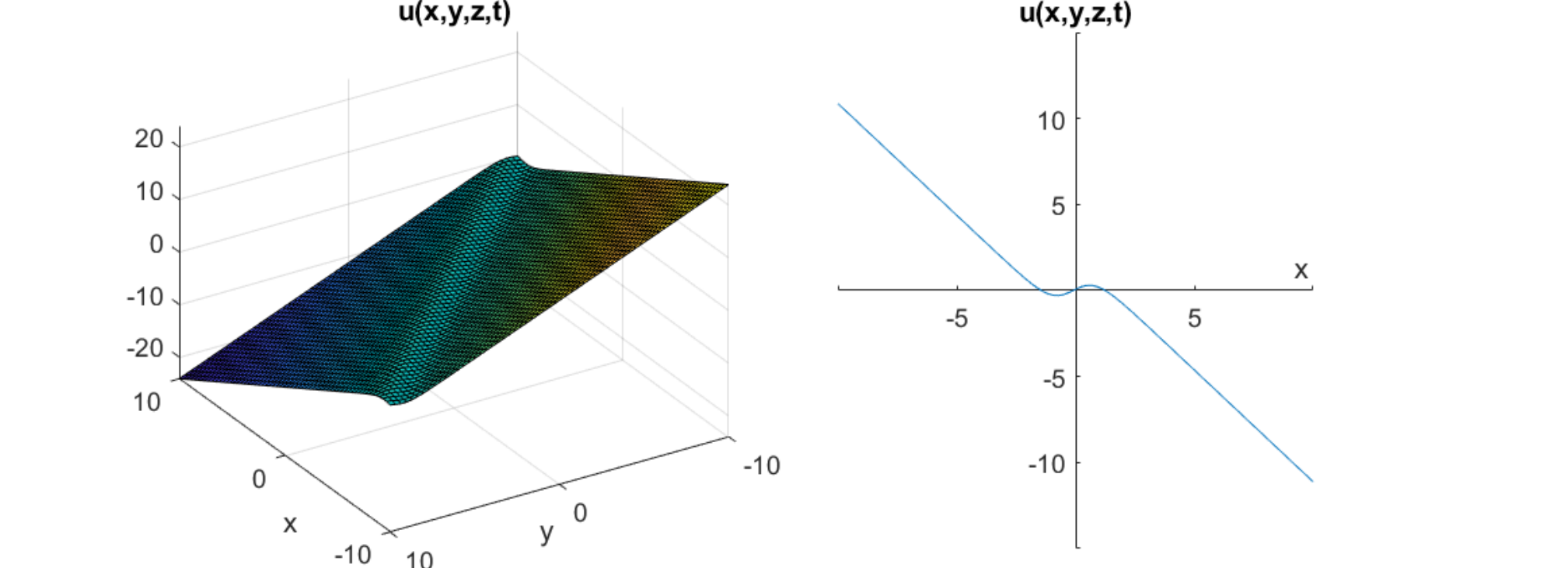}
    \caption{The graphical representation of Eq. \eqref{sol2} (3D on the left side and 2D on the right side). }
    \label{fig2}
\end{figure}
Case-3
\begin{equation*}
    c=-1\,,\,A_0=1\,,\,A_1=0\,,\,B_1=0\,,\,A_2=-1 \,,\,B_2=\pm i\,, 
    \end{equation*}
which gives:
\begin{equation}\label{sol3}
u(x,y,z,t)=\,U(\eta)=\, tanh\,(\eta) \pm\,i\,sech\, (\eta),\,\,\text{with}\,\,\eta=\, x+y+z+ t.
\end{equation} 
\begin{figure}[H]
\centering
    \includegraphics[trim={0.5cm 8.5cm 0cm 8.5cm},clip=true,totalheight=.35\textheight ]{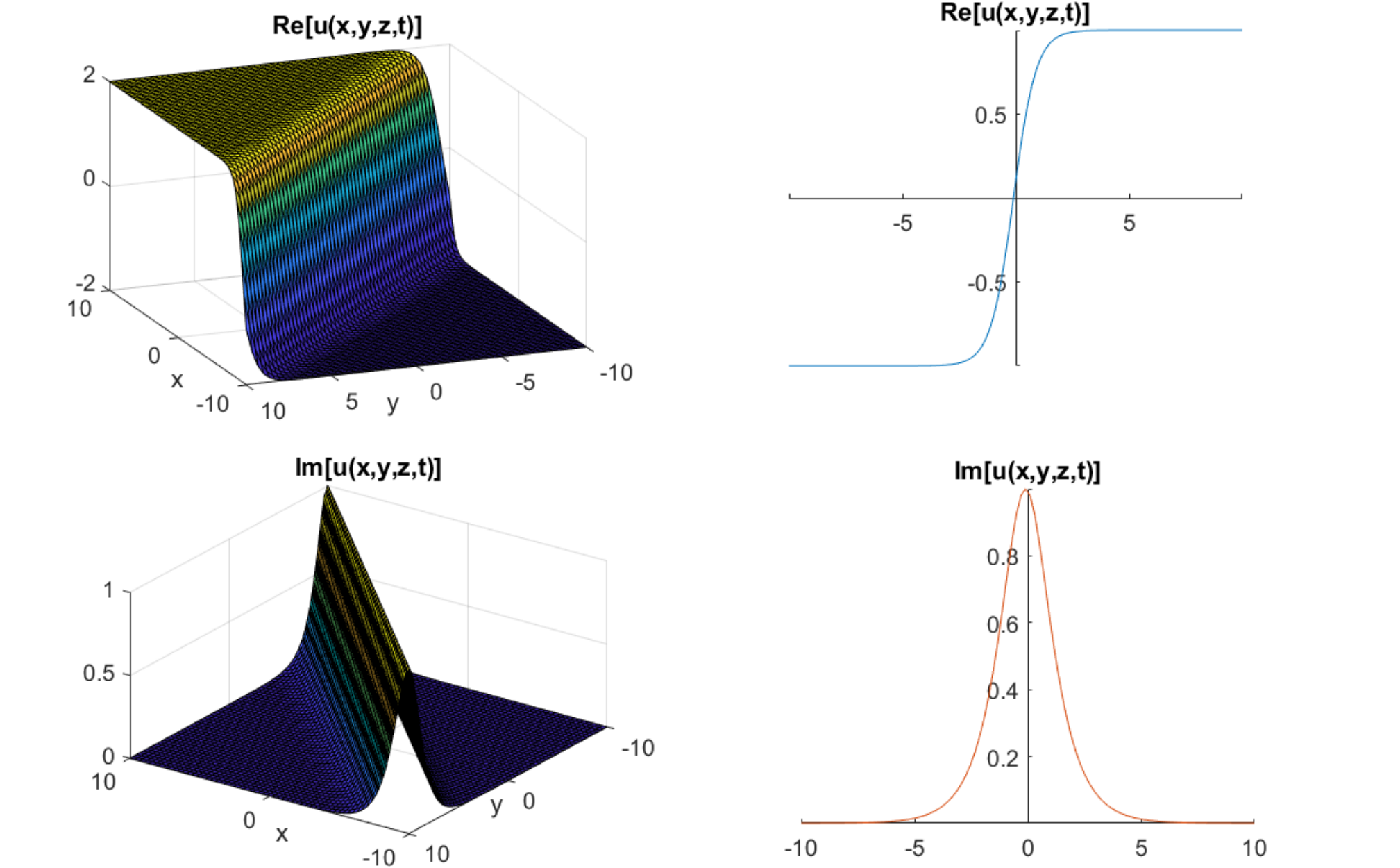}
    \caption{The graphical representation of Eq. \eqref{sol3} (3D on the left side and 2D on the right side).}
    \label{fig3}
\end{figure}
Case-4
\begin{equation*}
    c=-1/2\,,\,A_0=2/3\,,\,A_1=0\,,\,B_1=0\,,\,A_2=-1 \,,\,B_2=\pm i\,, 
    \end{equation*}
which gives:
\begin{equation}\label{sol4}
 u(x,y,z,t)\,=U(\eta)\,=tanh\,(\eta) \pm\,i\,sech\, (\eta)-\frac{1}{3}\,\eta, \,\, \text{with}\,\, \eta=x+y+z+\frac{1}{2} t.
\end{equation} 
\begin{figure}[H]
\centering
    \includegraphics[trim={0.5cm 8.1cm 0cm 8.1cm},clip=true,totalheight=.4\textheight ]{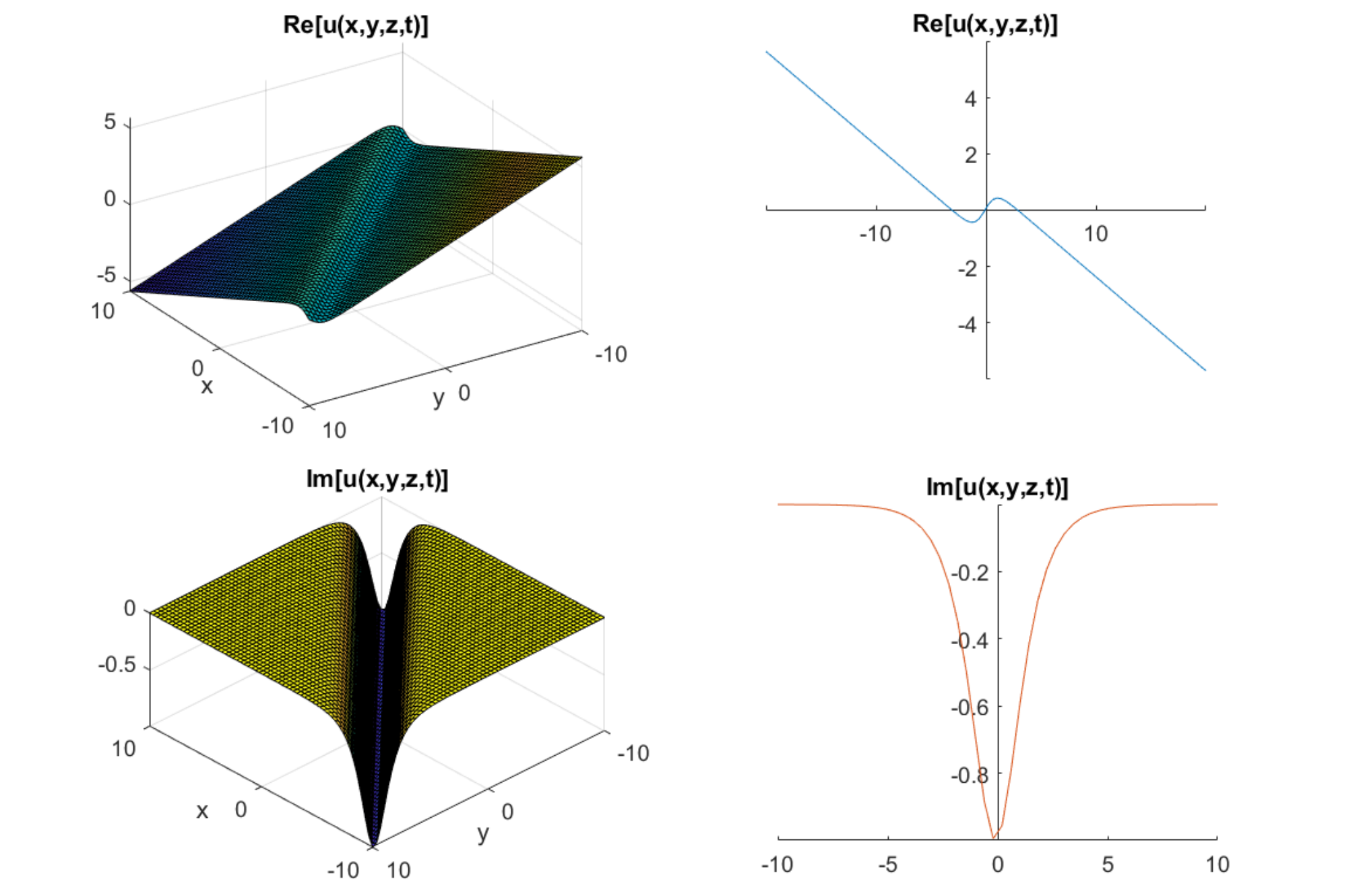}
    \caption{The graphical representation of Eq. \eqref{sol4} (3D on the left side and 2D on the right side).}
    \label{fig4}
\end{figure}
\subsection{The Reaction-Diffusion Equation}
We consider the following form of the reaction-diffusion equation \cite{a38}, 
\begin{equation}\label{eq20}
    u_{tt}+\alpha\, u_{xx}+\beta\, u+\gamma\, u^3=0\,
\end{equation}
where $\alpha$, $\beta$ and $\gamma$ are constants (nonzero). The moving coordinate $u(x,t)=U(\eta)$, where $\eta=x-ct$ leads the Eq. \eqref{eq20} in an ordinary differential equation (ODE) 
\begin{equation}\label{eq21}
    (\alpha+c^2)\, U''+\beta\,U+\gamma \, U^3=0.
\end{equation}
Considering homogeneous balance between $U''$ and $U^3$ in Eq. \eqref{eq21} based on Eq. \eqref{eq8} we obtain $n+2=n+n+n$, which implies $n=1$. Now we can write Eq. \eqref{eq8} as
\begin{equation}\label{eq22}
     U(\omega)= B_1\, sin (\omega)+A_1\, cos(\omega)+A_0.
    \end{equation}
Using Eq. \eqref{eq22} into Eq. \eqref{eq21}, and equaling all polynomials with same degree to get, we have a system of equation as below.\newline\\  
\newline
\newline
constant: \,\,\,\,\,\,\,\,\,\,\,\,\,\,$\gamma\,{A_{{0}}}^{3}+3\,\gamma\, A_{{0}}{B_
{{1}}}^{2}+\beta\,A_{{0}}
=0$\\
$cos(\omega):\,\,\,\,\,\,\,\,\,\,\,\,\,\,\,\,\,\,\,  3
\,\gamma\, {A_{{0}}}^{2}A_{{1}}+3\,\gamma\,A_{{1}}{B_{{1}}}^{2}-2\,{v}^{2}A_{{1}}-2\,
\alpha\,A_{{1}}+\beta\,A_{{1}} 
=0$\\
$sin(\omega):  \,\,\,\,\,\,\,\,\,\,\,\,\,\,\,\,\, \,\,        3\,\gamma\,{A_{{0}}}^{2}B_{{1}}+\gamma\,{B_{{1}}}^{3}-{v}^{2}B_{{1}}-\alpha\,B_{{1}}+
\beta\,B_{{1}}
=0$\\
$cos^2(\omega):\,\,\,\,\,\,\,\,\,\,\,\,\,\,\,\,\, 3\,\gamma\,A_
{{0}}{A_{{1}}}^{2}-3\,\gamma\,A_{{0}}{B_{{1}}}^{2}
=0$\\
$sin(\omega)\,cos(\omega):\,\,6\,\gamma\,A_{{0}}A_{{1}}B_{{1}}=
0$\\
$cos^3(\omega):\,\,\,\,\,\,\,\,\,\,\,\,\,\,\,\,\,\, \gamma\,{A_{{1}}}^{3}-3\,\gamma\,A_{{1}}{B_{{1}}}^{
2}+2\,{v}^{2}A_{{1}}+2\,\alpha\,A_{{1}}
=0$\\
$sin(\omega)\,cos^2(\omega):3\,\gamma\,{A_{{1}}}^{2}B_{{1}}-\gamma\,{B_{{1}}}^{3}+2\,{v}^{2}B_{{1}}+2\,
\alpha\,B_{{1}}
=0$\\
After solving the above system , we find the traveling wave solution $u(x,y,z,t)$ to Eq. \eqref{eq20}  in the form of \eqref{eq7}.
\par
Case-1
\begin{equation*}
    c=\pm\, \sqrt{\frac{\beta-2\,\alpha}{2}}\,,\,A_0=0\,,\,A_1=\pm\, \sqrt{\frac{-\beta}{\gamma}}\,,\,B_1=0\,,\alpha=\alpha\,,\,\beta=\beta \,,\,\gamma=\gamma, 
    \end{equation*}
which gives:

\begin{equation}\label{sol21}
 u(x,t)=U(\eta)=\pm\, \sqrt{\frac{-\beta}{\gamma}}\,tanh\,(\eta), \,\text{with}\,\, \eta=x\pm\, \sqrt{\frac{\beta-2\,\alpha}{2}} t.
\end{equation}
\begin{figure}[H]
\centering
    \includegraphics[trim={0.5cm 11cm 0cm 10cm},clip=true,totalheight=.25\textheight ]{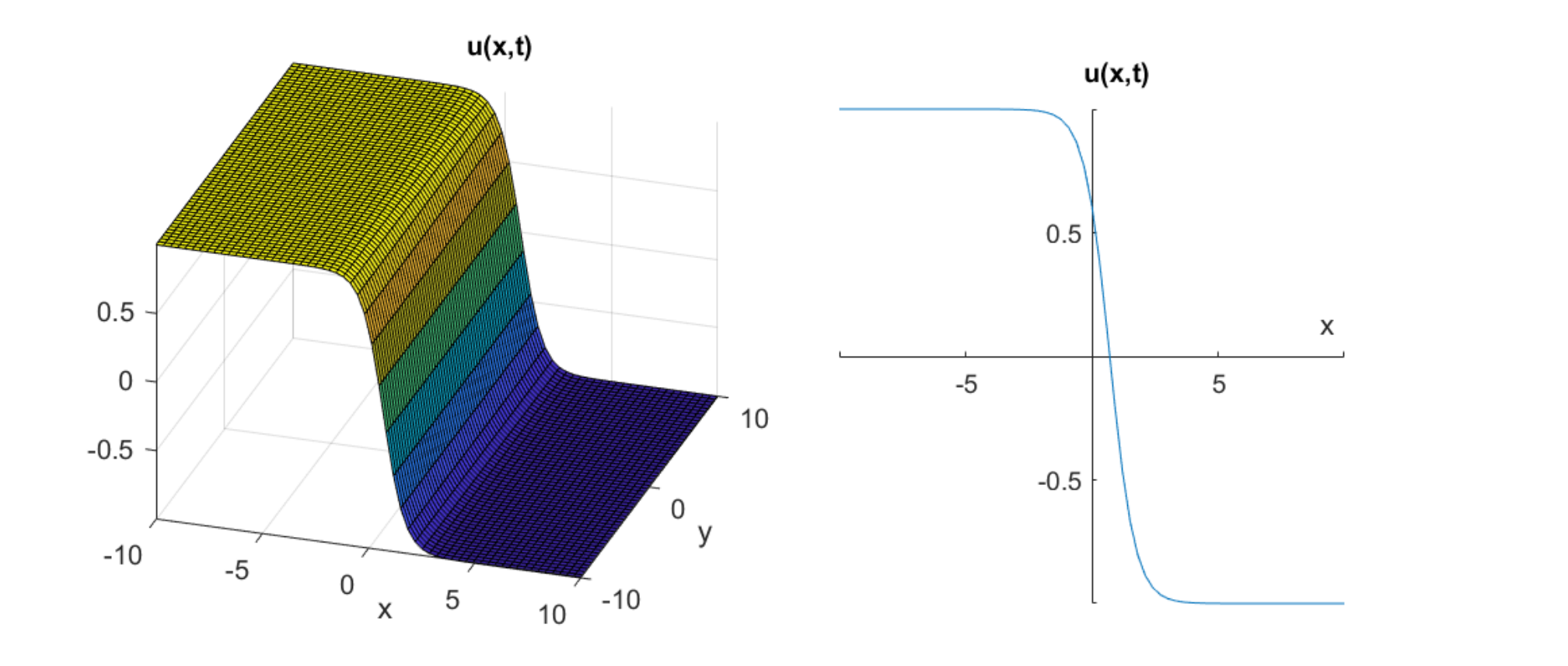}
    \caption{The graphical representation of Eq. \eqref{sol21} (3D on the left side and 2D on the right side).}
    \label{fig5}
\end{figure}
Case-2
\begin{equation*}
    c=\pm\, \sqrt{-\beta-\alpha}\,,\,A_0=0\,,\,B_1=\pm\, \sqrt{\frac{-2\,\beta}{\gamma}}\,,\,A_1=0\,,\alpha=\alpha\,,\,\beta=\beta \,,\,\gamma=\gamma,  
    \end{equation*}
which gives:    
\begin{equation}\label{sol22}
 u(x,t)=U(\eta)=\pm\, \sqrt{\frac{-2\,\beta}{\gamma}}\,sech\,(\eta),\,\, \text{with}\,\, \eta=x\pm\, \sqrt{-\beta-\alpha} t. 
\end{equation}
\begin{figure}[H]
\centering
    \includegraphics[trim={0.5cm 11.0cm 0cm 10.7cm},clip=true,totalheight=.25\textheight ]{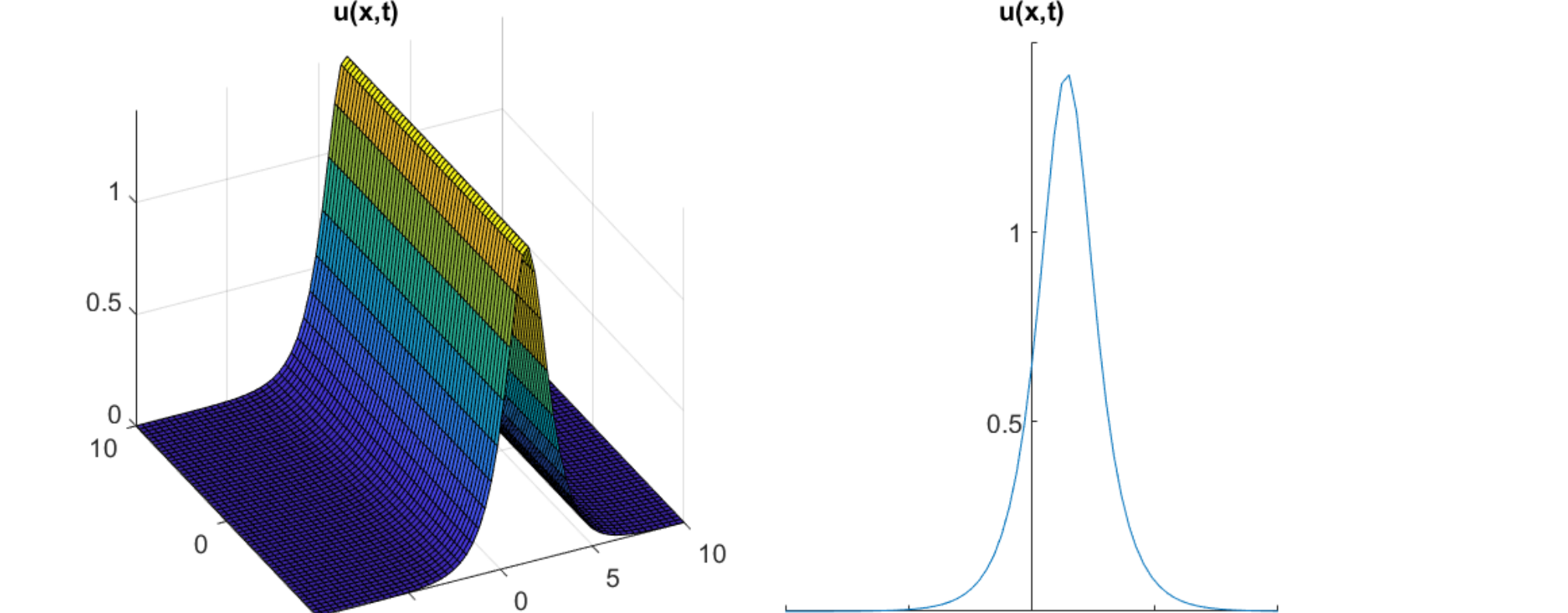}
    \caption{The graphical representation of Eq. \eqref{sol22} (3D on the left side and 2D on the right side).}
    \label{fig6}
\end{figure}
Case-3
\begin{equation*}
    c=\pm\, \sqrt{2\,\beta-\alpha}\,,\,A_0=0\,,\,A_1=\pm\, \sqrt{\frac{-\beta}{\gamma}}\,,\,B_1=\pm\, \sqrt{\frac{\beta}{\gamma}}\,,\alpha=\alpha\,,\,\beta=\beta \,,\,\gamma=\gamma, 
    \end{equation*}
which gives:    
\begin{equation}\label{sol23}
 u(x,t)=U(\eta)=\pm\, \sqrt{\frac{-\beta}{\gamma}}\,tanh\,(\eta)\pm\, \sqrt{\frac{\beta}{\gamma}}\,sech\,(\eta), \,\,\text{with}\,\,\eta=x\pm\, \sqrt{2\,\beta-\alpha} t.  
\end{equation}
\begin{figure}[H]
\centering
    \includegraphics[trim={1.0cm 8.2cm 2.5cm 8.5cm},clip=true,totalheight=.38\textheight ]{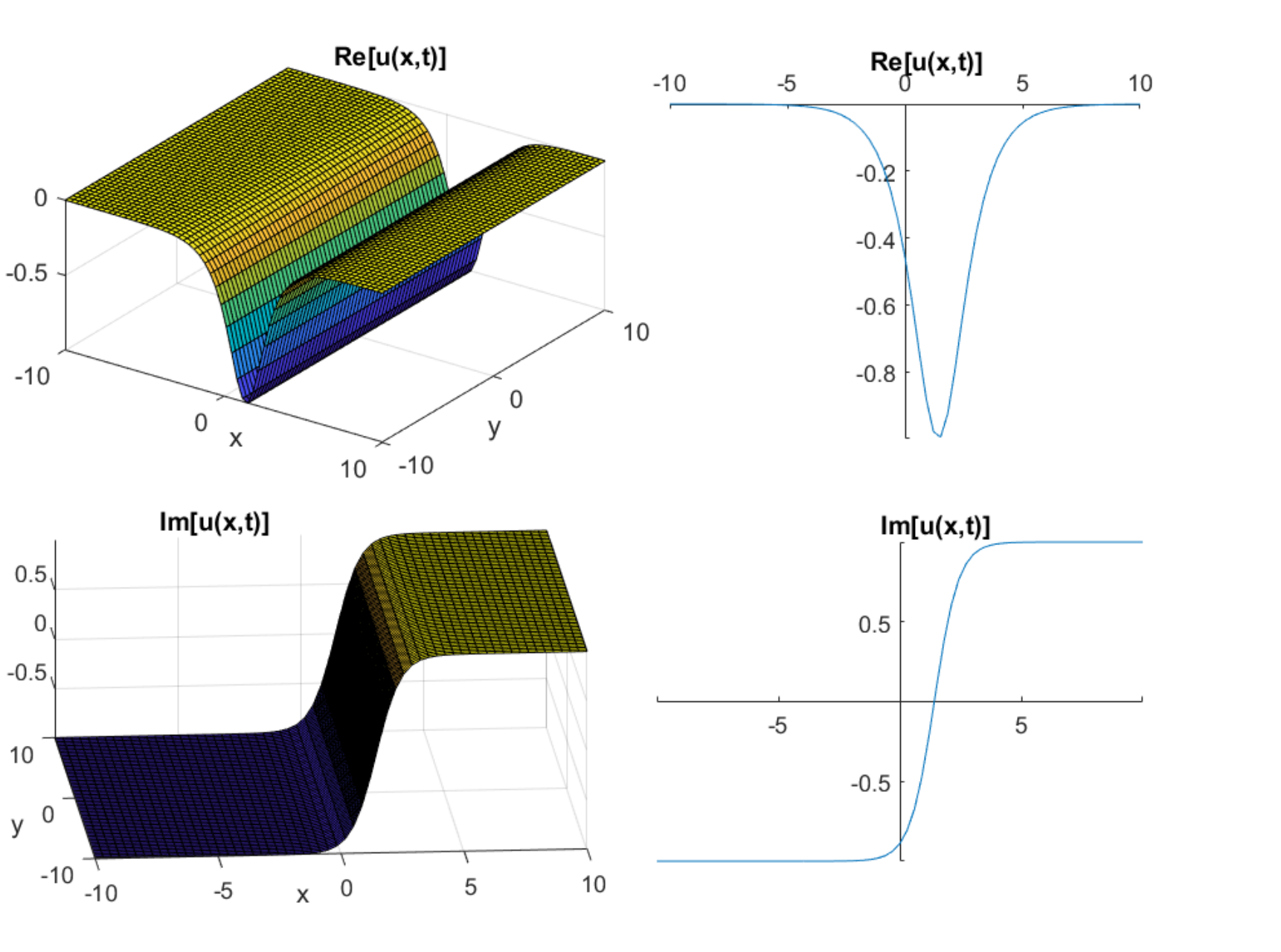}
    \caption{The graphical representation of Eq. \eqref{sol23} (3D on the left side and 2D on the right side).}
    \label{fig7}
\end{figure}

\section{Conclusion}
In this article, we have applied the Sine-Gordon expansion method for calculating new travelling wave solutions to the potential-YTSF equation of dimension (3+1) and the reaction-diffusion equation. We have found these solutions of the equation in the trigonometric, complex and hyperbolic function forms. This method is powerful and very efficient to finding travelling wave solutions to the NLEEs. We can solve various NLEEs by this method using any symbolic software.\newline
\newline 
{\Large \textbf{Abbreviations}} 
\newline
\newline
NLEEs, SGE and ODE. 
\newline
\newline
{\Large \textbf{Declarations}}\newline
\newline
{\Large \textbf{Availability of data and material}}: Not applicable. \\
\newline
{\Large \textbf{Competing interests}}:
The author has no any financial or non-financial conflict of interests.\\
\newline
{\Large \textbf{Funding }}: There was no funding for this research.\\
\newline
{\large \textbf{Authors' contributions}}: The Author did the research (methodology, examples and writing )  by himself and approved the final manuscript.  \\
\newline
{\Large \textbf{Acknowledgements}}: The author is very grateful to the editorial team and reviewers for their valuable comments and suggestions towards improving this article.

\bibliographystyle{IEEEtran}
\bibliography{sample}

\end{document}